\documentclass[useAMS,usenatbib,usegraphicx]{mn2e}

\title[Hydrogen gas in the edge-on dwarf galaxy UGC 1281]{Warp or lag?  The ionized and neutral hydrogen gas in the edge-on dwarf galaxy UGC 1281}

\author[P. Kamphuis et al.]{P. Kamphuis$^{1,2,3}$\thanks{E-mail:peter.kamphuis@astro.rub.de} , R.F. Peletier$^{1}$, P.C. van
der Kruit$^{1}$ and G.H. Heald$^{4}$\\
$^{1}$Kapteyn Astronomical Institute,University of Groningen, Postbus 800, 9700 AV,Groningen, the Netherlands \\
$^{2}$Astronomisches Institut der Ruhr-Universit\"at Bochum, Universit\"atsstr. 150, 44780 Bochum, Germany\\
$^{3}$Humboldt Research Fellow\\
$^{4}$Netherlands Institute for Radio Astronomy (ASTRON), 7990 AA Dwingeloo, the Netherlands }
\begin{document}

\date{}

\pagerange{\pageref{firstpage}--\pageref{lastpage}} \pubyear{2010}
\maketitle

\label{firstpage}
\begin{abstract}
The properties of gas in the halos of galaxies constrain global models of the interstellar
medium. Kinematical information is of particular interest since it is a clue to the origin of the gas. Until now mostly massive galaxies have been investigated for their halo properties.\\
\hspace*{0.5 cm}Here we report on deep {\sc H\,i} and H$\alpha$ observations of the edge-on dwarf galaxy UGC 1281 in order to determine the existence of extra-planar gas and the kinematics of this galaxy. This is the first time a dwarf galaxy is investigated for its gaseous halo characteristics. We have obtained H$\alpha$ integral field spectroscopy using PPAK at Calar Alto  and deep {\sc H\,i} observations with the WSRT of this edge-on dwarf galaxy. These observations are compared to 3D models in order to determine the distribution of {\sc H\,i} in the galaxy.\\
\hspace*{0.5cm}We find that UGC 1281 has  H$\alpha$ emission up to 25\arcsec (655 pc) in projection above the plane and in general a low H$\alpha$ flux.  Compared to other dwarf galaxies UGC 1281 is a normal dwarf galaxy with a slowly rising rotation curve that flattens off at 60 km s$^{-1}$ and a central depression in its {\sc H\,i} distribution. Its {\sc H\,i} extends 70\arcsec  (1.8 kpc) in projection from the plane. This gas can be explained by either a warp partially in the line-of-sight or a purely edge-on warp with rotational velocities that decline with a vertical gradient of 10.6 $\pm$ 3.7 km s$^{-1}$ kpc$^{-1}$. The line-of-sight warp model is the preferred model as it is conceptually simpler.  In either model the warp starts well within the optical radius.
\end{abstract}
\begin{keywords}
UGC 1281, Galaxies: halos, Galaxies: ISM, Galaxies: spiral, Galaxies: structure, Galaxies: kinematics and dynamics, Galaxies: dwarf
\end{keywords}

\section{Introduction}\label{intro}
The discovery of a very extended {\sc H\,i} halo in NGC 891 has shown that 21 cm observations are a powerful tool to study the structure and kinematics of gaseous halos. Since this discovery \citep{1997ApJ...491..140S} many investigations have followed studying this \citep{2007AJ....134.1019O} and other galaxies. Investigations at 21 cm, as well as other wavelengths, have shown that NGC 891 is not a special case in having a gaseous halo  \citep{2000A&A...356L..49S,2001A&A...377..759L,2003A&A...406..505R,2005A&A...439..947B,2005A&A...436..101W,2008A&A...490..555B}.\\
\hspace*{0.5 cm} It is thought that a large part of this halo gas is brought up from the disk by galactic fountains  \citep{1976ApJ...205..762S,1980ApJ...236..577B}. In this model supernovae explosions expel the hot hydrogen gas from the disk into the halo. In the halo the  gas cools down and condenses before returning to the disk. Studies of massive galaxies have shown that this mechanism is very likely at work in several galaxies, either inferred from the fact that the extra-planar {\sc H\,i} is concentrated towards the disk \citep{2007AJ....134.1019O} or that the majority of high velocity clouds are located near the bright inner disk \citep{2008A&A...490..555B}.\\
\hspace*{0.5 cm} The gas in the halo of NGC 891 shows rotational velocities that are lower  then the velocities of the gas in the disk. This 'lag' is observed in {\sc H\,i} \citep{2005ASPC..331..239F} as well as in H$\alpha$ \citep{2006ApJ...647.1018H,2007A&A...468..951K} and determined to be linear with a magnitude of  $\sim$16 km s$^{-1}$ kpc$^{-1}$ in the vertical direction.\\
\hspace*{0.5 cm}Simulations have shown that the lag cannot be explained by galactic fountains alone \citep{2008MNRAS.386..935F}. Together with other observational facts, such as  the low metallicity of the High Velocity Clouds (HVCs) and the presence of filaments and other irregular gaseous structures in the halo \citep{2008A&ARv..15..189S}, this has given rise to the idea that the halo must also be partially formed from gas that has been accreted from the Intergalactic Medium (IGM) or companion galaxies.  However, the gradients by themselves do not necessarily require accretion from the IGM \citep{2002ASPC..276..201B, 2006A&A...446...61B}.\\
\hspace*{0.5 cm}Even though this lagging behavior is observed in other galaxies then NGC 891  --e.g. \cite{2007ApJ...663..933H} studied two other large galaxies and found a lag in both of them-- it is not known whether it occurs in all gaseous halos. Until now, kinematical studies have focussed on massive galaxies  and one Low Surface Brightness (LSB)  galaxy  \citep{2003ApJ...593..721M} but dwarf galaxies have not yet been investigated for a kinematic lag. \\
\hspace*{0.5 cm}Here we present deep {\sc H\,i} and H$\alpha$ observations of the  edge-on  dwarf galaxy UGC 1281 and analyze the kinematics of the {\sc H\,i} and H$\alpha$ in and above the plane. The 21 cm line emission was observed with the Westerbork Synthesis Radio Telescope (WSRT)  and the  H$\alpha$ with the Integral Field Unit (IFU) PPAK on the 3.5 m telescope at Calar Alto.\\
\hspace*{0.5 cm}The edge-on orientation of UGC 1281 provides us with an excellent opportunity  to study the vertical structure of gas in a dwarf galaxy. Our {\sc H\,i} observations of UGC 1281 are some of the deepest  observations ever conducted on a dwarf galaxy, and never before has an edge-on dwarf galaxy been observed to this depth.\\
\hspace*{0.5 cm} UGC 1281 is a nearby edge-on dwarf galaxy (M$_{\rm{B}}$=-15.8, \cite{2002A&A...390..863S}) with a systemic velocity of 156 km s$^{-1}$ at a distance of 5.4 Mpc \citep{2004AJ....127.2031K}.  The galaxy has a angular diameter of  4.46\arcmin (D$_{25}$) \citep{1992yCat.7137....0D} on the sky which would translate to  7 kpc at the given distance. \cite{2000AJ....119.2757V,2001AJ....121.2003V} measured the  star formation rate in UGC 1281 to be very low (0.0084 M$_{\odot}$ yr.$^{-1}$).  However, this should be considered as a lower limit since the H$\alpha$ flux was not corrected for internal extinction. These observations are supported by the non-detections in radio continuum \citep{1991A&AS...87..309H} and IRAS. Indeed, \cite{2003A&A...406..505R} observed UGC 1281 as a part of their extra-planar diffuse ionized gas (DIG) survey  and could not detect any extra-planar H$\alpha$ in this dwarf galaxy. All these observations indicate a very low star formation rate in UGC 1281. \\
\hspace*{0.5 cm}Under the assumption that gaseous halos are created by processes related to star formation, little to no halo is expected when the SFR is low. However, in a dwarf galaxy gas might easily escape from the disk due to the weak potential. Also, even though small galaxies are not expected to accrete baryonic matter at lower redshifts \citep{2010arXiv1001.4721H}, UGC 1281 is in the transition region, between accreting and non-accreting galaxies, and therefore a modest amount of lagging extra-planar gas might be present if the vertical gradient is predominantly formed by the accretion of matter. \\ 
\hspace*{0.5 cm}This article is structured as followed. In $\S$ \ref{obs}  we will describe the observations and data reduction. $\S$ \ref{results} will contain the results of the observations followed by a presentation of our models in $\S$ \ref{models}. We will discuss our results in $\S$ \ref{disc} and summarize in $\S$ \ref{summary}.
 \section{Observations \& Data Reduction}\label{obs}
 \subsection{Radio Data}
The 21 cm line emission, or {\sc H\,i}, observations were obtained with the WSRT during four nights in September 2004. In total 4 complete 12 hr observations were performed using the Maxi-Short configuration.  This configuration gives the optimum performance for imaging extended sources.  This is because it provides a good sampling of the inner $U-V$ plane with a shortest base line of 36 m. The longest  baseline is 2754 m. An overview of the observational parameters is given in Table \ref{obsparHI}. \\
\begin{table}
\begin{center}
\begin{tabular}{ll}
\hline Parameter&Value\\ 
\hline 
Observation date & 2004 September\\
Total length of observation (hr) & 4 $\times$ 12\\
Velocity center of band (km s$^{-1}$) & 156\\
Total bandwidth (MHz) & 10\\
Channels in obs. & 1024\\
Channel sep. in obs. (kHz) & 9.77 \\
Channels in final cube & 61\\
Vel. res. after Hanning smoothing (km s$^{-1}$) & 8.24 \\
\hline
\end{tabular}
\caption{Log of the {\sc H\,i} observations.} \label{obsparHI}
\end{center}
\end{table}
\hspace*{0.5 cm}The data  were reduced using the MIRIAD package \citep{1995ASPC...77..433S}. Care was taken not to include data affected by solar  or other interference. Before and after each 12 hr observation a calibration source (3C147 and CTD93) was observed, thus enabling us to determine the spectral response of the telescope. During each 12 hr observation no additional (phase) calibration sources were observed, as is standard practice with the  WSRT. Due to the large bandwidth,  the time variations can be determined from self-calibration of detected sources in the continuum image constructed from the channels free of line emission. In this case UGC 1281 itself was not detected in continuum emission.\\
\hspace*{0.5 cm}After the reduction and calibration additional analysis was performed with the GIPSY package \citep{1992ASPC...25..131V}. The final cube was reduced to 61 velocity channels with a velocity spacing of 4.12 km s$^{-1} $, which results in a velocity resolution after Hanning smoothing of 8.24 km s $^{-1}$. The original cube has a spatial resolution of 25.0\arcsec $\times$ 13.4\arcsec. This cube was smoothed to two cubes with circular beams of 26\arcsec (0.68 kpc) and 60\arcsec  (1.57 kpc) to avoid beam orientation effects and to detect low level emission (see Table \ref{cubeparHI}). \\
\hspace*{0.5 cm}For the calculation of the minimum detectable column in Table \ref{cubeparHI} we used a velocity width of 16.5 km s$^{-1}$, because real emission will never appear in a single channel.\\ 
\begin{table*}
\begin{center}
\begin{tabular}{lccc}
\hline Parameter&Full Resolution& Circular Beam & Low Resolution\\ 
\hline 
Spatial resolution (\arcsec)&25.0 $\times$ 13.4&26.0 $\times$ 26.0&60.0 $\times$ 60.0\\
Beam size (kpc)&0.65 $\times$ 0.35&0.68 $\times$ 0.68&1.57 $\times $1.57\\
rms noise per channel (mJy beam$^{-1}$)&0.44&0.50&0.72\\
Minimum detectable column density (3$\sigma$; cm$^{-2}$)&7.1 $\times$ 10$^{19}$ &4.0 $\times$ 10$^{19}$&1.1 $\times$ 10$^{19}$\\
\hline
\end{tabular}
\caption{Parameters of the {\sc H\,i} Data Cubes.} \label{cubeparHI}
\end{center}
\end{table*}
The position of the center of the galaxy, as determined by \cite{2002A&A...390..863S} from $R$-band photometry, was set to zero in the three cubes and they were rotated by 50 $^\circ$ (PA = 40$^\circ$, \cite{2002A&A...390..863S}) to orient the major axis of the galaxy parallel to the x-axis of the image.  \\
\subsection{IFU data}
The H$\alpha$ was observed with the PPAK integral field unit on the 3.5 m telescope at Calar Alto \citep{2006PASP..118..129K}. Three positions were observed in one night (see Table \ref{obsparHa} ). At each position several exposures were taken.  Two positions were observed for 3 $\times$ 1200 s (North and Center) and one for 3 $\times$ 600 s (South). The lower exposure time of the Southern field was due to twilight. The pointings of the North and South fields were shifted by $\sim \pm$ 76\arcsec, compared to the Center,  along an axis with a PA slightly offset ($\sim 10^{\circ}$ offset) from the PA of the galaxy. This was done to ensure a good coverage of extra-planar H$\alpha$. The spectra cover a wavelength range $\sim$5500 - 7000 \AA \hspace{0.1 cm}  with a resolution of 4.1 \AA  (187 km s$^{-1}$ at H$\alpha$). The conditions were partially photometric with a mean seeing of 1.3\arcsec, which is much less than the fiber size (2.7\arcsec). \\
 \begin{table}
\begin{center}
\begin{tabular}{lll}
\hline Parameter&Field 1\&2  & Field 3\\ 
\hline 
Name & North\&Center & South\\
Observation date & 2006 September  & same\\
Exposure time (s) & 3600  & 1800\\
Central wavelength (\AA) & 6273  & 6273\\
Total bandwidth (\AA) & 1628 & 1628 \\
Channels in obs. & 1050  &1050\\
Channel sep. in obs. (\AA) & 1.55 & 1.55 \\
Channel separation (km s$^{-1}$) at H$\alpha$ & 70.8 & 70.8 \\

\hline
\end{tabular}
\caption{Log of the H$\alpha$ observations. } \label{obsparHa}
\end{center}
\end{table}
\hspace*{0.5 cm}For the reduction of the PPAK spectra the IRAF package was used. The steps in the {\tt dohydra} Guide \citep{iraf-guide} were followed manually to ensure complete control over the reduction. 
The only deviation taken from the steps as described in this guide was that  we traced the apertures on the science frames themselves. This was done because there was more than enough continuum emission in each spectrum and there was no need to introduce additional errors  due to shifting the apertures.   \\
\hspace*{0.5 cm}For the initial wavelength calibration 7 lines of a HeNe+ThAr lamp were used. Afterwards a  fine tuning calibration was performed with 5 sky lines in each science frame. For the sky line subtraction a total of 95 sky fibers from the three pointings were used.  All sky fibers were checked for inconsistencies and, except for sky fibers with the galaxy in their field of view, none were found. \\
\hspace*{0.5 cm}After the reduction and wavelength calibration the different exposures of each field were combined by calculating the median of the separate exposures. No clipping was applied. The three fields were then positioned into one field by overlaying the continuum images of the fields on top of a $R$-band DSS  image. By taking special care that the five stars in the three fields were aligned properly, this procedure resulted in a position error  less than 1\arcsec, which is  smaller than the 2.7\arcsec\hspace{0.1cm}fiber size.\\
\hspace*{0.5 cm}Our  reduced and calibrated data were further analyzed  with a combination of  IDL programs produced by the SAURON collaboration (e.g. \cite{2002MNRAS.329..513D}) and ourselves.  The data were Voronoi binned with the voronoi\_2d\_binning IDL program \citep{2003MNRAS.342..345C} to obtain a higher S/N ratio (S/N $>$  20). Like the {\sc H\,i}, the central position of the galaxy was set to (0,0) and the galaxy was rotated by 50$^\circ$ to align the major axis with the x-axis.
\section{Results}\label{results}
Here, we  first discuss the distribution and kinematics of the hydrogen that can be directly derived from the data. In Section \ref{disc} we will compare the distributions and kinematics of the ionized and neutral gas, highlight the differences and similarities, and discuss possible interpretations of these distributions. 
 \subsection{Gas distribution}\label{dist}
 \subsubsection{H$\alpha$ Distribution}\label{Hadist}
  \begin{figure*}
\centering
\includegraphics[angle=0,width=14cm]{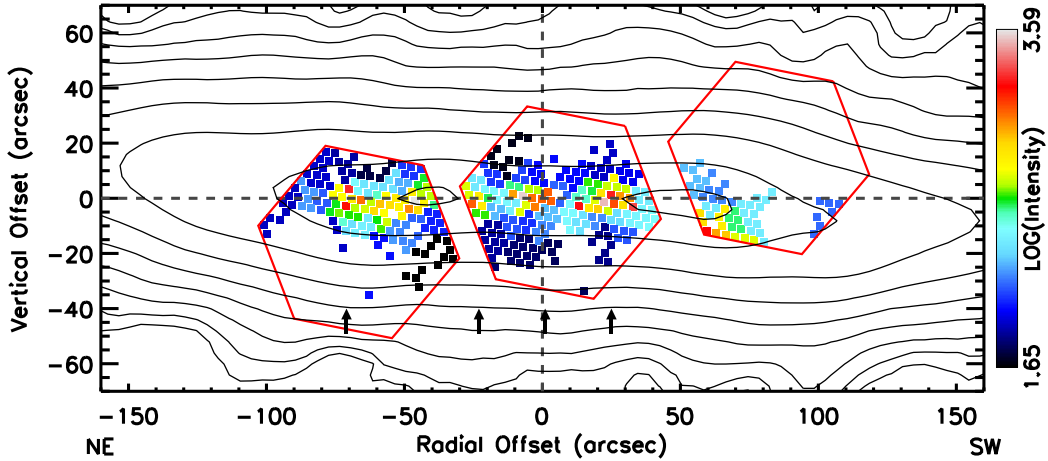}
\caption{Integrated H$\alpha$ flux map of UGC 1281. The map was constructed by taking the log of the area under the fitted Gaussian in each spectrum. The red lines indicate the full field of view of the PPAK observations. Overlaid in black are the contours of an integrated {\sc H\,i} (zeroth moment) map. The contours are at 1.5, 3, 6, 12, 24, 48, 96, 192, 273 $\sigma$ levels of the data where $\sigma=2.2$ mJy beam$^{-1}$.  The arrows indicate the positions of the velocity cuts parallel to the minor axis in Fig. \ref{vertrotcurves}.} \label{mom0Ha}
\end{figure*}
Fig. \ref{mom0Ha} shows the H$\alpha$ flux obtained with the PPAK IFU instrument.  The flux in each bin was determined by fitting a Gaussian, with the IDL routine GAUSSFIT, to an average of the spectra in each bin. The area under this Gaussian is then considered to be the flux in each bin. All the fitted Gaussians were inspected by eye, and none of the fitted lines showed significant deviations from Gaussian. For display purposes the log values of the flux are displayed. For the same reason the bins without emission are simply taken away from the data and thus not displayed. Overlaid on the image are the contours of an integrated {\sc H\,i}  (zeroth moment) map. \\
\hspace*{0.5 cm}The first thing we see from Fig.  \ref{mom0Ha} is that the ionized hydrogen is mostly located in 5 or 6 distinct peaks but that there is low level emission almost everywhere in the central field of view. The distinct peaks can most likely be associated with large {\sc H\,i\,i} regions in the galaxy whereas the low level emission is indicating a diffuse ionized component.  The two innermost peaks at the NE and SW side of the center of the galaxy might be caused by a ring-like structure or central depression in the distribution. This seems to be implied by their equal distance from the center; however, they could also be normal {\sc H\,i\,i} regions.\\
\hspace*{0.5 cm} A warp is clearly seen in the {\sc H\,i\,i} regions as well as the diffuse emission. The peak of the emission starts to deviate from the major axis at a radial distance of $\sim$ 50\arcsec (1.3 kpc). In the South West (SW) of the galaxy we see an exception to this behavior with a large  ionized hydrogen peak below the major axis, where the diffuse emission still seems to be mostly on the major axis.  This seems to indicate an {\sc H\,i\,i} region that is either somewhat offset from the plane or located in the outskirts of the galaxy. If the warp is partially along the line of sight the outer parts will not only experience a change in position angle but also in inclination. Thus if this {\sc H\,i\,i} region is located in the outskirts of the galaxy its position can be in the (warped) plane of the galaxy.\\
\hspace*{0.5 cm} The H$\alpha$ distribution could, in theory, be severely affected by internal dust  extinction, especially in the edge-on orientation. However, in the case of UGC 1281 this seems unlikely because no clear dust lane can be observed in HST imaging of UGC 1281 \citep{2008ASPC..388..395B}  and dwarf galaxies are expected to have a low metallicity \citep{2004A&A...423..427P}  and therefore little dust content. The reddish color  of UGC 1281 \citep{1998A&AS..133..181M} is most likely caused by the absence of a large young stellar population. This once more confirms the idea that UGC 1281 has a low SFR. \\
\hspace*{0.5 cm} In our data  the maximum  distance to the mid-plane where diffuse gas is still detected is $\sim$25\arcsec (0.65 kpc). This is similar to the the extent of the stars ($\frac{1}{2}d_{25,minor}=18$\arcsec,  \cite{2000AJ....119.2757V}, $\frac{1}{2}d_{25,minor}=23$\arcsec, \cite{1992yCat.7137....0D} ).  If we fit an exponential to the inner vertical intensity profile of the ionized gas we find a scale height of 8.5\arcsec (0.22 $\pm$ 0.03 kpc) assuming that the galaxy is seen perfectly edge-on. When we follow the same procedure for the continuum emission in our spectra and an $I$-band image\footnote{obtained through the NASA extragalactic database} we find a scale height of 7.6\arcsec (0.20 $\pm$ 0.01 kpc) and 8.0\arcsec (0.21 $\pm$ 0.01 kpc) respectively.\\
\subsubsection{{\sc H\,i} Distribution}\label{HIdist}
\begin{figure*}
\centering
\includegraphics[angle=0,width=14cm]{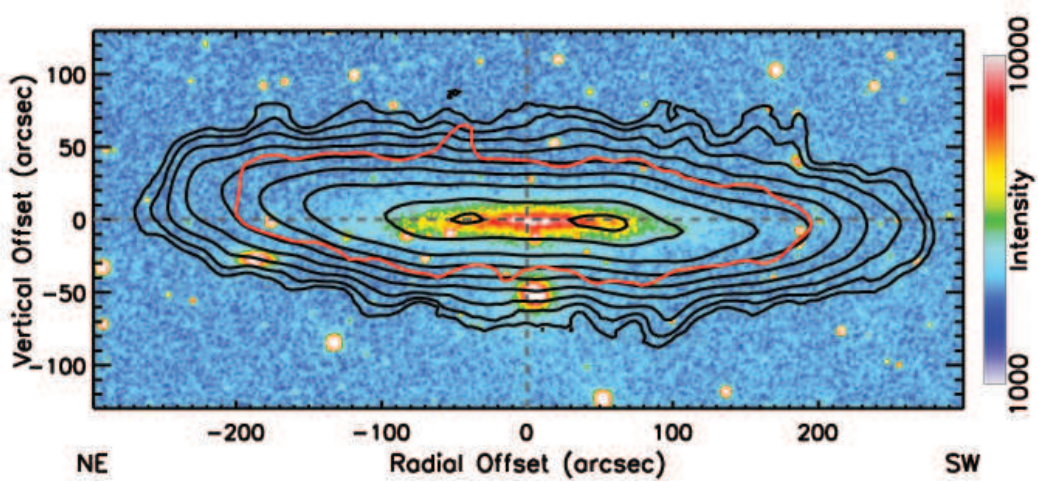}
\caption{Continuum image in the red from the DSS 2 overlaid with contours of our integrated  {\sc H\,i} flux map of UGC 1281 (see text). The black contours are the same as in Fig. \ref{mom0Ha}. The red contour indicates the 3$\sigma$ contour of the WHISP observations.} \label{mom0HI}
\end{figure*}
\begin{figure}
\centering
\includegraphics[angle=0,width=8cm]{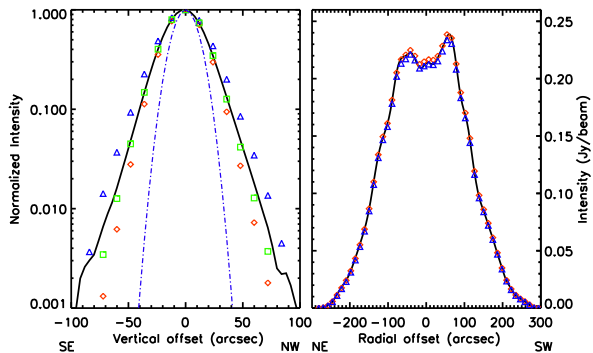}
\caption{Intensity profiles of the observed {\sc H\,i}. Left figure: Vertical intensity profile, averaged over the inner 100\arcsec in the radial direction and normalized to the peak intensity.  Right figure: Radial intensity profile, averaged over the inner 40\arcsec of the galaxy in the vertical direction. Black line: Data, Green squares: Best fit model, Red diamonds (Blue triangles): best fit model with a scale height -2\arcsec (+4\arcsec) (see $\S$ \ref{models}). Blue dotted dashed line: beam. } \label{intprof}
\end{figure}

Fig. \ref{mom0HI} shows the DSS 2  red image of UGC 1281 overlaid with the contours of our integrated {\sc H\,i} flux map. This {\sc H\,i} map was constructed by  adding all the channels of the Circular Beam data cube. To keep the addition of noise to a minimum, only regions that  are above 3$\sigma$ in the Low Resolution cube (see Table \ref{cubeparHI}) were considered.\\
\hspace*{0.5 cm}From Fig. \ref{mom0HI} we can see that the {\sc H\,i} of UGC 1281 is at first glance quite symmetrically and evenly distributed.  However, a closer look reveals asymmetries and peculiarities in the {\sc H\,i} distribution. It warps away from the major axis at about  90\arcsec (2.4 kpc) on the South West and at about  100\arcsec (2.6 kpc)  on the North East. The warp initially shows the normal S-shape observed in many edge-on galaxies \citep{2002A&A...394..769G} but bends back to the plane of the inner disk at larger radii. This behaviour is seen especially at the South West side, at a radial offset of $\sim$200\arcsec. 
This warp was already observed by \cite{2002A&A...394..769G} in the WHISP observations of this galaxy \citep{2001ASPC..240..451V}.\\
\hspace*{0.5 cm} When we compare the lowest contour  in the integrated moment map of the WHISP observations with our own (see Fig. \ref{mom0HI}, red contour), we see that in our observations more emission is detected in the radial as well as in the vertical direction.  The growth in both directions is similar in extent and this indicates that even with our deep observations we might not be detecting the lowest levels of emission of this galaxy.\\
\hspace*{0.5 cm} Furthermore, the {\sc H\,i} distribution displays a central depression. This depression appears  as a ring around the center of the galaxy and ranges from  10\arcsec to 40\arcsec (0.26-1.05 kpc) radial offset from the center of the galaxy. It appears to be symmetrical around the center of the galaxy. \\
\hspace*{0.5 cm}The {\sc H\,i} in UGC 1281 shows significant extensions away from the major axis. The {\sc H\,i} extends up to  70\arcsec (1.8 kpc) on both sides of the plane at column densities $N_{HI}=4.0\times 10^{19} \rm{cm}^{-2} (3\sigma)$. This extent is much more than the FWHM of the beam (26\arcsec) which is clearly seen in Fig. \ref{intprof}. This figure shows the vertical distribution of the data (black solid line) and a Gaussian (blue dot-dashed line) with a FWHM of 26\arcsec. Both are normalized to the maximum of the data. In this figure it is easily observed that the wings of the data are much more extended than the observational beam. \\
\subsection{Velocity Distribution}\label{Vel}
\subsubsection{H$\alpha$ Velocities}\label{Havel}
\begin{figure*}
\centering
\includegraphics[angle=0,width=14cm]{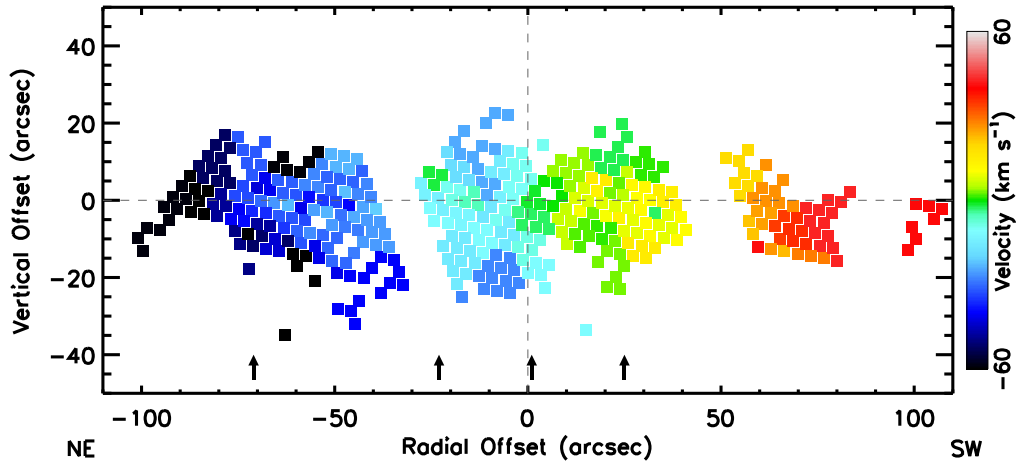}
\caption{Velocity field of the ionized gas. The field was constructed by taking the central position of the fitted Gaussian in all binned spectra. The systemic velocity ($\mathrm{V_{sys}}=156.$ km s$^{-1}$) has been set to 0. The separate pixels show the fiber positions and the colors run from -60 to 60 km s$^{-1}$. The arrows indicate the positions of the velocity cuts parallel to the minor axis in Fig. \ref{vertrotcurves}.} \label{mom1Ha}
\end{figure*}
\begin{figure*}
\centering
\includegraphics[angle=0,width=17cm]{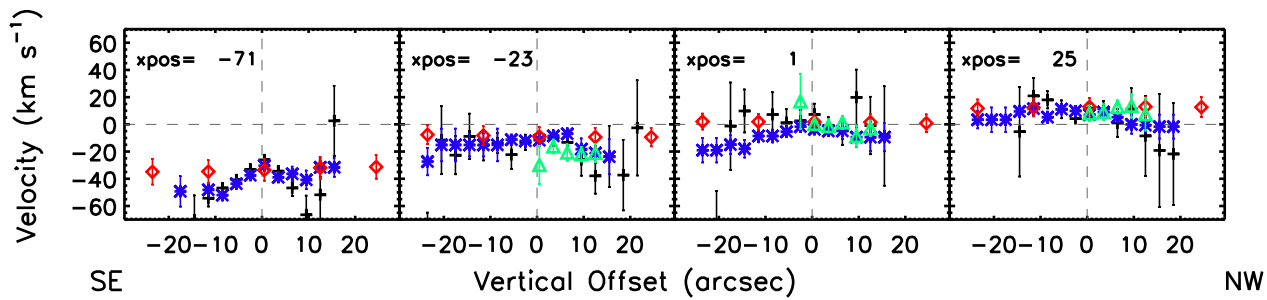}
\caption{Four cuts parallel to the minor axis through the velocity field at  -71, -23, 1, 25\arcsec radial offset from the center. The negative (positive) offsets are the South East (North West) in the sky.  Black points are the unbinned PPAK data. Blue, voronoi binned PPAK data. Red, {\sc H\,i} data. Green, \citet{2006ApJS..165..461K} data.} \label{vertrotcurves}
\end{figure*}

Fig. \ref{mom1Ha} shows the velocity field of the PPAK observations. This velocity field was obtained by taking the peak position of the Gaussian fitted to each bin (see $\S$ \ref{Hadist}). This is by no means  equal to the actual deprojected maximum rotational velocity in the galaxy but it is an apparent mean velocity determined by a combination of the rotational velocity, the density distribution of the ionized gas, and the opacity of the dust. From here on whenever we mention velocity we are referring to this mean velocity unless otherwise noted. The gaussian fitting procedure, and therefore the mean velocity, was chosen because with a channel separation of 70 km s$^{-1}$ it is impossible to confidently fit  the intrinsic shape of the emission line. \\
\begin{figure}
\centering
\includegraphics[angle=0,width=8 cm]{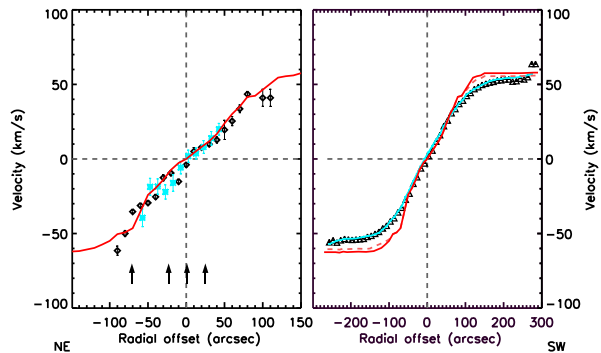}
\caption{Velocities along the major axis. The left panel shows the H$\alpha$  velocities: Black symbols are the PPAK data presented in this paper, blue symbols are the data obtained by \citet{2006ApJS..165..461K}.  Right figure shows {\sc H\,i}  velocities: Black symbols are the {\sc H\,i},  the blue line shows the velocities obtained from the model. The red lines are the input  unprojected rotation curves of the best fit models described in $\S$ \ref{models}. The arrows indicate the positions of the velocity cuts parallel to the minor axis in Fig. \ref{vertrotcurves}.} \label{rotcurves}
\end{figure}
\hspace*{0.5 cm}Fig. \ref{rotcurves} (left) shows a cut 10\arcsec wide of the velocity field along the major axis. Overplotted are the velocities obtained by \cite{2006ApJS..165..461K} with the  DensePAK IFU (blue symbols) and the rotation curve obtained from the modeling (see $\S$ \ref{models}). \cite{2006ApJS..165..461K} were not able to trace emission as far out in radius with their observations. Since their exposure time and fiber size is equal to ours  this is most likely caused by the fact that they do not bin the data and the lower sensitivity of the DensePak IFU. The velocities obtained by \cite{2006ApJS..165..461K} agree well with our values, which assures us that there are no systematic errors in our reduction or the gaussian fitting procedure.\\ 
\hspace*{0.5 cm}In this plot we see clearly that the part of the galaxy observed in our H$\alpha$ field of view is still resembling a slow rising rotation curve that indicates solid body rotation. This behavior of the rotation curve is quite typical for dwarf galaxies, which all seem to have a large inner region in which their rotation curve resembles solid body rotation \citep{2000AJ....120.3027C}. This behavior could be caused by us considering the mean velocities. However, this is unlikely  as we will discuss in $\S$ \ref{HIvel}.  \\
\hspace*{0.5 cm}Another thing that is quite clear  is that the flux peaks in the H$\alpha$ along the major axis stand out in the velocity curve  as areas with lower velocities (Fig. \ref{rotcurves}, left, black arrows). This is expected if they are {\sc H\,i\,i} regions, since the chance that they would lie on the line of nodes is low, and would imply that most of the peaks we see in the H$\alpha$ distribution are indeed {\sc H\,i\,i} regions. Another possible explanation is  that the intensity peaks correspond to a higher density in the radial density profile of the galaxy with thicker clumps offset from the line perpendicular to the line of sight. This explanation is supported by the fact that some of the peaks appear symmetrically around the center of the galaxy (see Section \ref{Hadist}) and that the star formation rate in UGC 1281 is so low (SFR = 0.0084 M$_{\odot}$ yr.$^{-1}$) that giant {\sc H\,i\,i} regions would not be expected to reside in the galaxy.\\
\hspace*{0.5 cm}If so, this would require that as the influence of the overdensity or {\sc H\,i\,i} region, on the shape of the line profile diminishes, the measured velocities should increase. We can see that this is the case when we look away from the mid-plane as is illustrated in Fig. \ref{vertrotcurves}. These plots show the  velocity as function of vertical offset from the plane in bins 12\arcsec wide for the binned and unbinned H$\alpha$ (blue and black, respectively), the data from \cite{2006ApJS..165..461K} (green), and the {\sc H\,i} (red) at the positions of the peaks.\footnote{Note that also above the major axis, the DensePak data and our data agree within the errors.} Here we see that the velocities move away from the systemic velocity as we look higher above the plane. This confirms the idea that the velocities on the major axis are lowered by a line-of-sight projection effect. These results and their implications will be discussed in Section \ref{Hap} in combination with the results of the {\sc H\,i}.
\subsubsection{{\sc H\,i} Velocities}\label{HIvel}
\begin{figure*}
\centering
\includegraphics[angle=0,width=16cm]{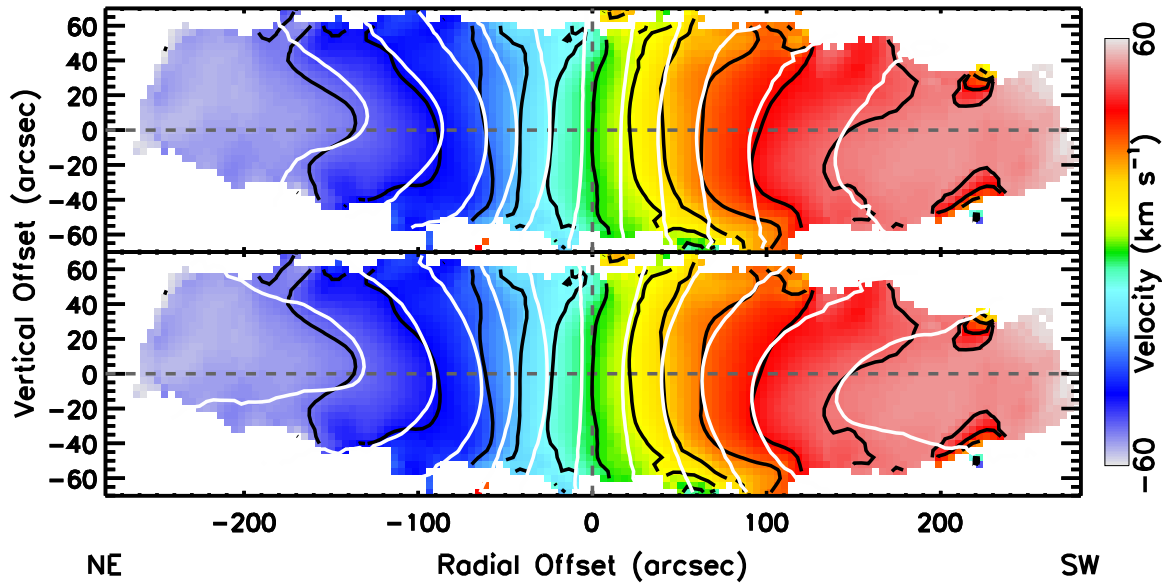}
\caption{Velocity field of the neutral hydrogen. The field was constructed with the task MOMENTS in GIPSY (see text). Contours are from -60 km s$^{-1}$ to 60 km s$^{-1}$ with increasing steps of 10 km s$^{-1}$. Black contours are the data. The white contours in the upper panel show the line-of-sight  warp model. In the lower panel the white contours show the best fit edge-on  model with a lag of 10.6 km s$^{-1}$ kpc$^{-1}$ (see $\S$ \ref{models}). } \label{mom1HI}
\end{figure*}

\begin{figure*}
\centering
\includegraphics[angle=0,width=16cm]{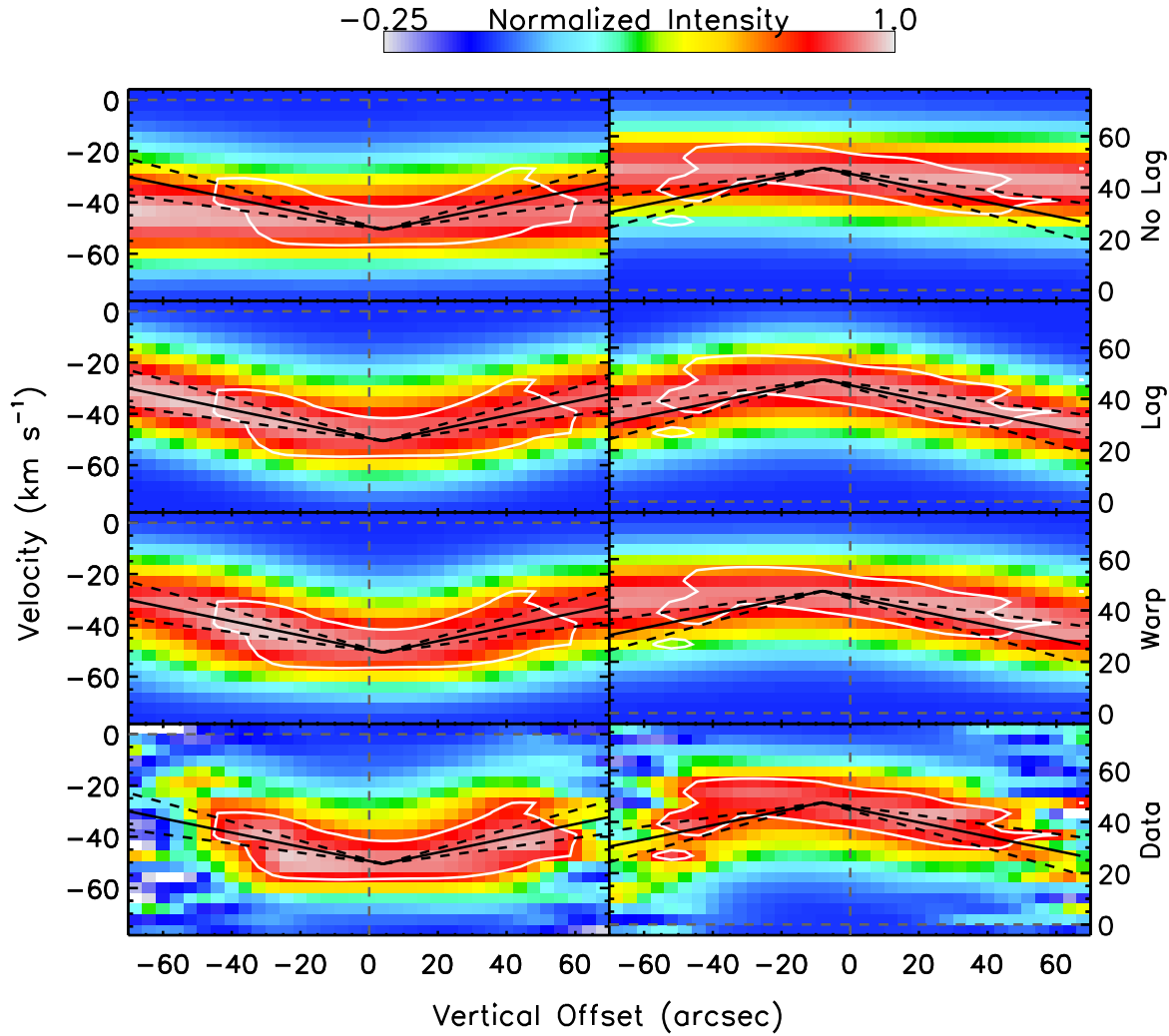}
\caption{Normalized PV diagrams parallel to the minor axis averaged over 100\arcsec at a radial offset of -175\arcsec to -75\arcsec (left) and 75\arcsec to 175\arcsec.  From top to bottom, the edge-on model without lag, the edge-on model with lag, the line-of-sight warp model and the data. The white contour outlines the 0.75 level of the data. The solid lines show the lag obtained from an edge-on lagging model (see $\S$ \ref{HImodsec}). The dashed lines show the derived error on the lag. The zero point is set to the point of maximum intensity in a non-normalized PV diagram.} \label{norm}
\end{figure*}

Fig. \ref{mom1HI} shows the velocity field of the {\sc H\,i} observations. This velocity field was constructed with the MOMENTS routine in GIPSY by selecting the first moment of the data cube. The routine determines the intensity weighted mean velocity position of the peak of the line profile in every pixel of the cube.  For symmetric profiles, this is analogous to fitting a gaussian profile to the emission line and taking its velocity at the peak. Since UGC 1281 has only small rotational values the line profiles are almost symmetric.  We have checked this by comparing a map with the velocities where the line profiles have their maximum with this GIPSY map and we find no differences greater than 6 km s$^{-1}$, which is less than the velocity resolution (see Table \ref{obsparHI}). Thus, these velocities are in principle comparable one to one with the velocities of the H$\alpha$ and differences should be due to the conditions of the gas (distribution effects, dust, resolution, self absorption, or a real difference in the rotational speed of the ionized and neutral gas).\\
\hspace*{0.5 cm}It is common to retrieve the rotation curve of edge-on galaxies through envelope tracing. Theoretically this is the correct way of retrieving the rotation curve of an edge-on galaxy \citep{1979A&A....74...73S}, however there are several reasons why Gaussian fitting is preferable in an edge-on dwarf system. As explained in the previous paragraph  the velocities in a dwarf galaxy are small, this means that the dispersion of the gas makes up a significant part of the line profile. Since in envelope tracing, either by fitting gaussians or  a scaling of the maximum intensity \citep{2001ARA&A..39..137S}, this dispersion is a chosen parameter it increases the uncertainty of envelope tracing. Furthermore, envelope tracing is also an estimation of the real rotational velocities and detailed modeling is still required. Therefore we prefer the well understood and described method of Gaussian fitting.  This does mean that only from the modeling we obtain information about the real rotation curve. \\ 
\hspace*{0.5 cm}Fig. \ref{rotcurves} (right) shows a cut 10\arcsec wide of the velocity field along the major axis. Here we see the same slow rise in the rotation curve as seen in the H$\alpha$ but we also see that in the {\sc H\,i} we do reach the flat part of the rotation curve at $\sim$120\arcsec radial offset, outside the H$\alpha$ field of view,  from the center of the galaxy at a maximum velocity $\sim$60 km s$^{-1}$.\\
\hspace*{0.5 cm}If we move away from the plane (Fig. \ref{mom1HI} and \ref{norm})  we see that the  velocities are lower than on the major axis. This implies that either the outer parts of the galaxy are heavily inclined, that the disk is flaring or that the gas above the plane is 'lagging'. 
We can quantify this vertical velocity gradient in normalized PV diagrams  (see \cite{2007A&A...468..951K}, $\S$ 6.2) parallel to the minor axis (Fig. \ref{norm}, bottom row) by fitting a straight line to the maxima between 20\arcsec and 50\arcsec  offset from the plane. Of course this fit is affected by the warp, lag, beam smearing and possibly other systematics. Therefore, the measured value has to be compared to the same measurements of models that can explain the observed declining velocities. These models will be presented in the next section and discussed in Section \ref{disc}.\\
\section{Models}\label{models}
The H$\alpha$ observations show such an irregular distribution that constructing a model for comparison  will not bring more insight into the structure of the H$\alpha$ distribution. Also, the distribution is peaked in many places whereas our models would resemble a smooth exponential distribution. We therefore decided to forego any attempt at modeling the ionized hydrogen in 3D. However, there are several parameters that can be derived from the data or obtained from the literature. These are presented  in Table \ref{Hapara} for comparison to the {\sc H\,i} model.
\begin{table}
\begin{center}
\begin{tabular}{lll}
\hline Parameter&Value& Reference\\ 
\hline 
EM Scale height & 0.22 $\pm$ 0.03 kpc& [1] \\
Scale length& 1.09 kpc&[2]\\
SFR & 0.0084 M$_{\odot}$ yr. $^{-1}$ &[2]\\
H$\alpha$ luminosity & 3.81$\pm$0.01 $\times 10^{-13}$ erg s$^{-1}$ cm$^{-2}$ &[3]\\
\hline
\end{tabular}
\caption{H$\alpha$ parameters that could be measured from the observations or obtained from the literature. [1] this paper, [2] \citet{2001AJ....121.2003V}, [3] \citet{2000AJ....119.2757V}. }\label{Hapara}
\end{center}
\end{table}
\subsection{Tilted ring models}\label{HImodsec}
\begin{figure*}
\centering
\includegraphics[angle=0,width=16cm]{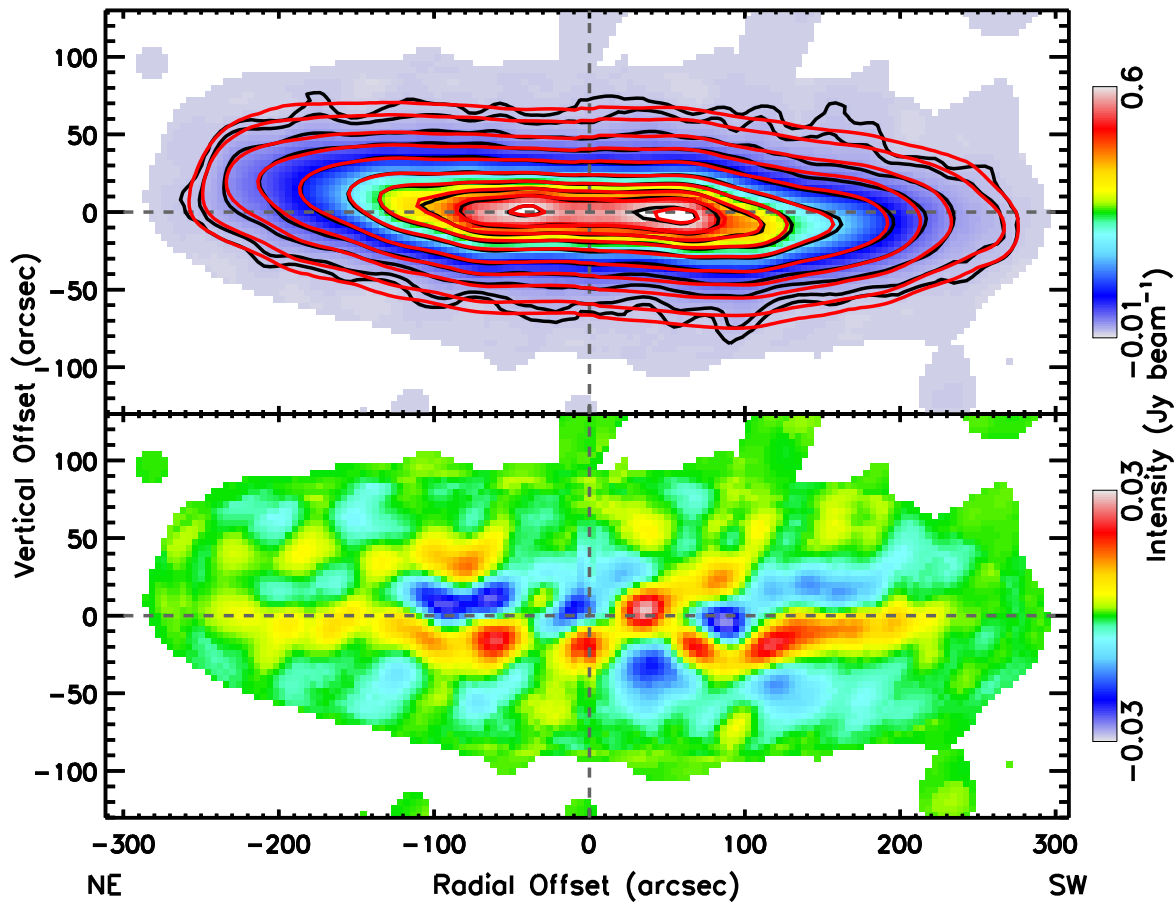}
\caption{Upper panel: Integrated {\sc H\,i} flux map of UGC 1281 (see Fig. \ref{mom0HI}). The black contours are the data at the same levels as in Fig. \ref{mom0Ha}. The red contours are the same levels for the best fit edge-on model. Lower panel: Residual map of the data minus the best fit edge-on model (total intensity).} \label{mom0HI2}
\end{figure*}
\begin{figure*}
\centering
\includegraphics[angle=0,width=16cm]{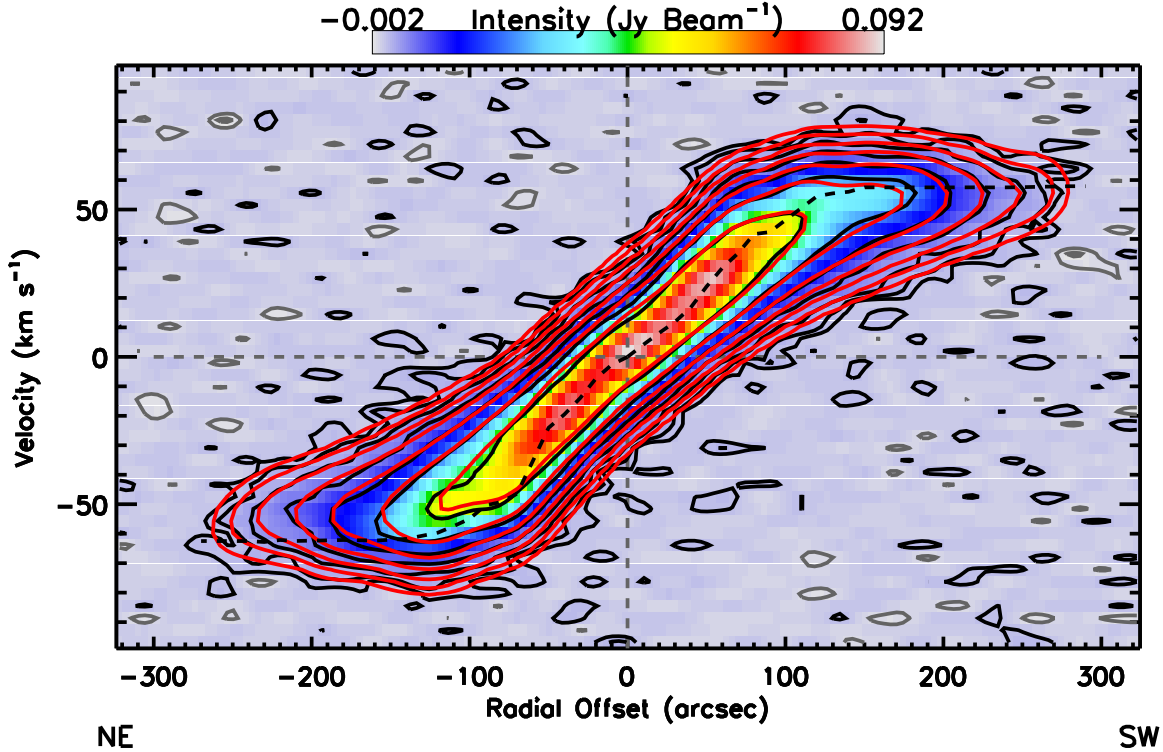}
\caption{Color plot of the {\sc H\,i} PV diagram along the major
axis. Overlaid with contours of the data (grey/black) and the best fit model (red). The contour levels are -3$\sigma$, -1.5$\sigma$, 1.5$\sigma$, 3$\sigma$ and 6$\sigma$  (1 $\sigma$ = 0.5 mJy/beam) etc. } \label{xv0}
\end{figure*}
\begin{figure}
\centering
\includegraphics[angle=0,width=8cm]{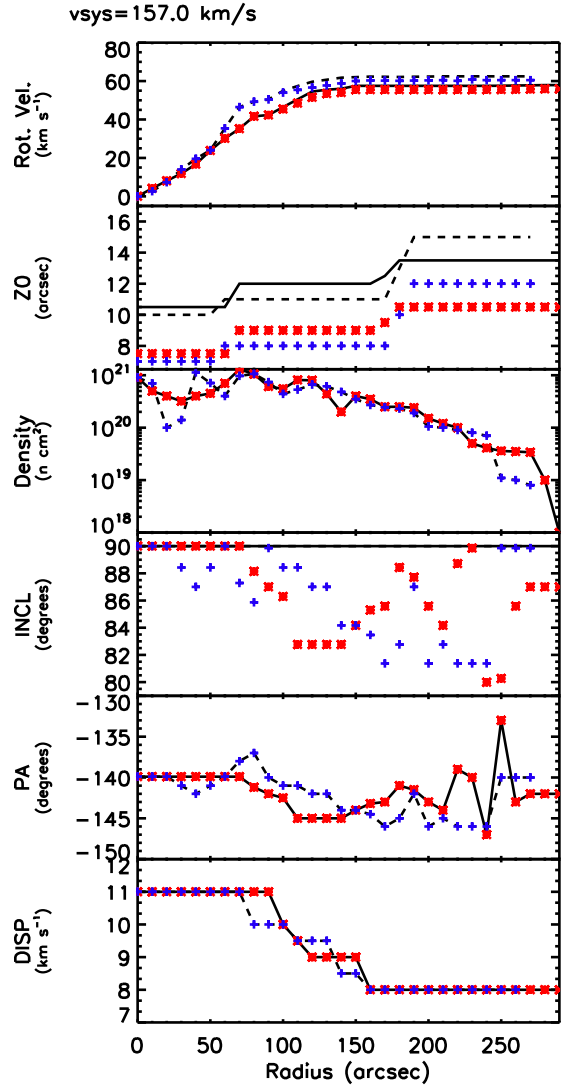}
\caption{Model parameters as function of radius. For the edge-on model (Solid line: South West side, Dashed line: North East side) and the maximum line-of-sight  warp model (Blue plusses: South West side, Red crosses: North East side).} \label{HIpara}
\end{figure}

To better understand the 3D distribution and the kinematics of the neutral hydrogen in UGC 1281 we have constructed a  range of tilted ring models. These models were created  with the GIPSY task  GALMOD. The GALMOD routine was modified by one of us (G. H.) and F. Fraternali to be able to include radial and vertical motions as well as a vertical gradient for the rotational velocities.\\
\hspace*{0.5 cm}The modeling is not straightforward, since many different parameters may be independently fitted and some of them are degenerate. In the case of an edge-on galaxy matters are even more complicated because small changes in density in the outer rings can seriously affect the inner rings of the model.  For this reason the fitting always has to start at the outer radii and then move inward ring by ring. Since it is clear that the warp is not symmetric at all, we fit both sides of the galaxy independently. The goodness of a fit was determined by investigating several representations of the data (PV diagrams, residual maps) by  eye.\\
\hspace*{0.5 cm} The following, iterative, procedure was followed for the fitting. We start by comparing radial and vertical density profiles of the data with the models (see Fig. \ref{intprof}). This has to be done to get a initial guess of the density profile and scale height.  When these profiles have a reasonable fit we start fitting the position angle on a integrated moment map. The position angle was deemed to fit when, at all radii, the vertical position of the peak value was  equal in the models and the data.\\
\hspace*{0.5 cm}In the tilted ring model the inclination of each ring is related to the position angle and dependent on the angle between the axis on which the ring tilts and the line of sight. We assume a simple warp, where the tilt axii for all rings are aligned, e.g. a straight line of nodes. The assumption of a straight line of nodes is a simplifying one that lowers the degrees of freedom in the model, making it possible to constrain the vertical distribution and kinematics in the model. By assuming a simple warp the inclination of each ring is coupled to its position angle by just one parameter, the angle between the line of sight and  the line of nodes. We shall refer to this angle as the warp axis angle.\\ 
\hspace*{0.5 cm}From the data it is impossible to determine an exact warp axis angle. We therefore set out to find the maximum and minimum warp axis angle that can fit the observations. Starting at a minimum warp axis angle of 0$^{\circ}$, e.g. all rings  90$^{\circ}$ inclined, we found an acceptable density distribution immediately.\\
\hspace*{0.5 cm}We then set out to find the maximum warp axis angle. This maximum is set at the point where the vertical profile starts to show a break, which is not observed in the data (see Fig. \ref{intprof}). By slowly increasing the warp axis angle and comparing the vertical distribution of the model to the data we find that this break becomes significant when the warp axis angle is greater than 60$^{\circ}$, thus defining the maximum line-of-sight warp model to have a warp axis angle less than 60$^{\circ}$.\\
\hspace*{0.5 cm}At this stage we split our modeling into two parts. Besides a model with a warp axis angle of 0$^{\circ}$, the edge-on model, we construct a second model with a warp axis angle of  55$^{\circ}$, the maximum line-of-sight  warp model.  Now the only parameter of the density that is not determined yet is the vertical distribution of each ring. In our models the vertical density, of all the rings, declines as an exponential of which the steepness is set by its scale height. By increasing/decreasing the scale height of our models systematically we find the upper and lower limit of the scale height. We allow the scale height to change from ring to ring but only increase toward larger radii. This increase would resemble a flare. In this way we find that the best fit scale height of the {\sc H\,i} in UGC 1281 is a flaring model that ranges from  10\arcsec to 15\arcsec (0.26 to 0.39 kpc) for our edge-on model and from 7\arcsec to 12\arcsec (0.18 to 0.31 kpc) for our maximum line-of-sight warp model.\\
\hspace*{0.5 cm} In order to estimate the accuracy of our scale heights we  systematically raise and lower the best fit flaring model. It turns out that the vertical profile significantly deviates from the observations when we add (subtract) more then 4\arcsec (2\arcsec) (100 and 50 pc respectively) to the scale heights of our best fit models. The vertical distributions of these models are shown in Fig. \ref{intprof}  as blue triangles (upper limit) and red diamonds  (lower limit) for the edge-on model. The blue dot-dashed line indicates the spatial resolution (FWHM = 26\arcsec).\\ 
\hspace*{0.5 cm}As a last check on the density distribution we construct an integrated moment map of the residual cube. Fig. \ref{mom0HI2} shows the residual map of the best fit density distribution. This figure shows only the density distribution of the best fit edge-on model because, above the sensitivity limit of the data, the differences in the density distribution between the two models are less than 5\%. Therefore the difference between these two models would not be visible in such plots. In this figure the wiggles that necessitate adding a warp, purely in PA, to the edge-on model, and their reproduction in this model, are clearly seen. \\
\hspace*{0.5 cm}  After we obtain a satisfactory density distribution we start fitting the rotation curve and dispersion of the gas. Initially we set the dispersion of the gas to be constant at 9 km s$^{-1}$ and then start fitting the rotation curve. As a first guess for a lower limit of the rotation curve we use the velocities from the H$\alpha$ and {\sc H\,i}. For the inner $\pm$ 50\arcsec we use the H$\alpha$  velocities and beyond this point we use the velocities obtained from the {\sc H\,i} first moment map (see Fig. \ref{rotcurves}). This rotation curve is then raised from the outside in until it fits the data.\\
\hspace*{0.5 cm} As an upper limit we start from a flat rotation curve and fit this curve by lowering the velocities from the inside out. The change between rings is always half of the difference between the previous two rings, with a minimal increase of 4 km s$^{-1}$ for the inner rings.  Because a low inner density could artificially lower the observed velocities in an edge-on system we have set the densities of the inner rings to zero while fitting this rotation curve.\\
 \hspace*{0.5 cm}Once more the fitting is an iterative process where the goodness of the fit is determined by comparing the major axis position-velocity (PV) diagram of the data  to the model. An example of such an PV diagram is shown in Fig. \ref{xv0}, where the best fit edge-on model is plotted with the data. The color scheme and the black contours are the data whereas the red contours display our best fit edge-on model. Again only the edge-on model is shown because also in this PV diagram the differences between the two models are less than 5\%. When we have obtained satisfactory fits to this PV diagram for both the upper and lower limit rotation curves  we find that they differ no more then a channel width (4.1 km s$^{-1}$) at any ring position which indicates that the rotation curve is well constrained.   Besides the upper and lower limit rotation curves matching very well, also the rotation curves of both, independently fitted, sides of the galaxy match up quite well. The final rotation curves are shown in Fig. \ref{rotcurves} and  \ref{HIpara} (Top Panel). These curves are the average of the upper and lower limit of both best fit models. Fig. \ref{xv0} show the rotation curve of the best fit edge-on model as a black dashed line overlaid on the major axis PV diagram.\\
\hspace*{0.5 cm}After fitting the rotation curve we found that the fit to the major axis PV diagram could be significantly improved by introducing a gradient in the velocity dispersion. This gradient runs from $\sigma_{\rm v}$=11 km s$^{-1}$ in the center to $\sigma_{\rm v}$=8 km s$^{-1}$ at the largest radii (see Fig. \ref{HIpara}, Bottom Panel). \\
\hspace*{0.5 cm} The model cubes should now be comparable to the data cube everywhere and any major deviations can only be caused by gas that is deviating from co-rotation at high projected distances from the mid-plane. In the case of the maximum line-of-sight  warp model the observations are fully reproduced by the best fit model. However, in the case of the edge-on model the velocities above the plane are clearly over estimated. This can be seen in Figure \ref{norm}. When we compare the edge-on model (Upper panels) to the data (Lower panels) it is easily seen that in the data the emission peaks at a vertical offset lie at lower velocities than the emission peak in the mid-plane. However, such a bending is only produced by the warp when the rings are not edge-on and thus completely lacking in the edge-on model without a lag. Therefore we reproduce the edge-on model with a rotation curve that declines as a function of distance to the plane, e.g. a lag. We construct seven models where the vertical gradient is increased from 4 to 28 km s$^{-1}$ kpc$^{-1}$ in steps of 4 km s$^{-1}$ kpc$^{-1}$.\\
\begin{figure}
\centering
\includegraphics[angle=0,width=8cm]{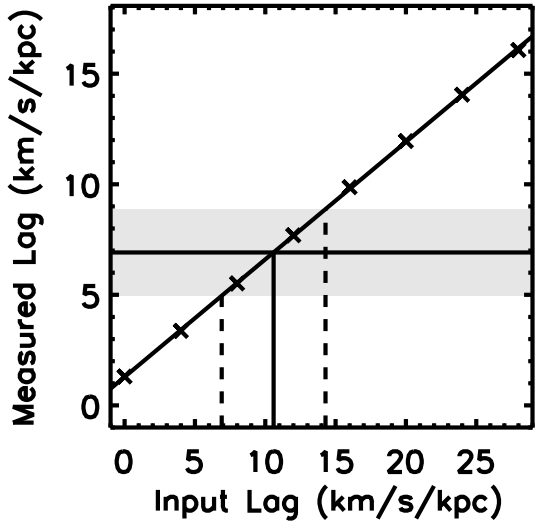}
\caption{Input lag vs. Measured lag in the models. The solid line shows the fit to the measured values (displayed by crosses) of the models. The horizontal solid line displays the value obtained from the data. The vertical solid line shows where the measured value intersects the model fit. The grey shaded area indicates the error in the measurement and the dashed vertical lines where the error intersects the fit. This range determines the error bar on the value of the lag.} \label{lag}
\end{figure}
\hspace*{0.5 cm} When we measure the vertical gradient in these models in the same way as in the data (see $\S$ \ref{HIvel}),  we find that the value measured from the data corresponds to a lag of 10.6 $\pm$ 3.7 km s$^{-1}$ kpc$^{-1}$ (see Fig. \ref{lag}) in the edge-on model. After adding this lag to the model it was found that, due to beam smearing effects, the major axis PV-diagram showed emission at slightly lower velocities than the data and the line-of-sight warp model. To correct for this effect the flat part of the input rotation curve was raised by 2 km s$^{-1}$. With this correction the major axis PV-diagrams of both models agreed within 5\% again. A model with this lag produces a satisfactory fit at all heights.   \\
\begin{figure*}
\centering
\includegraphics[angle=0,width=8cm]{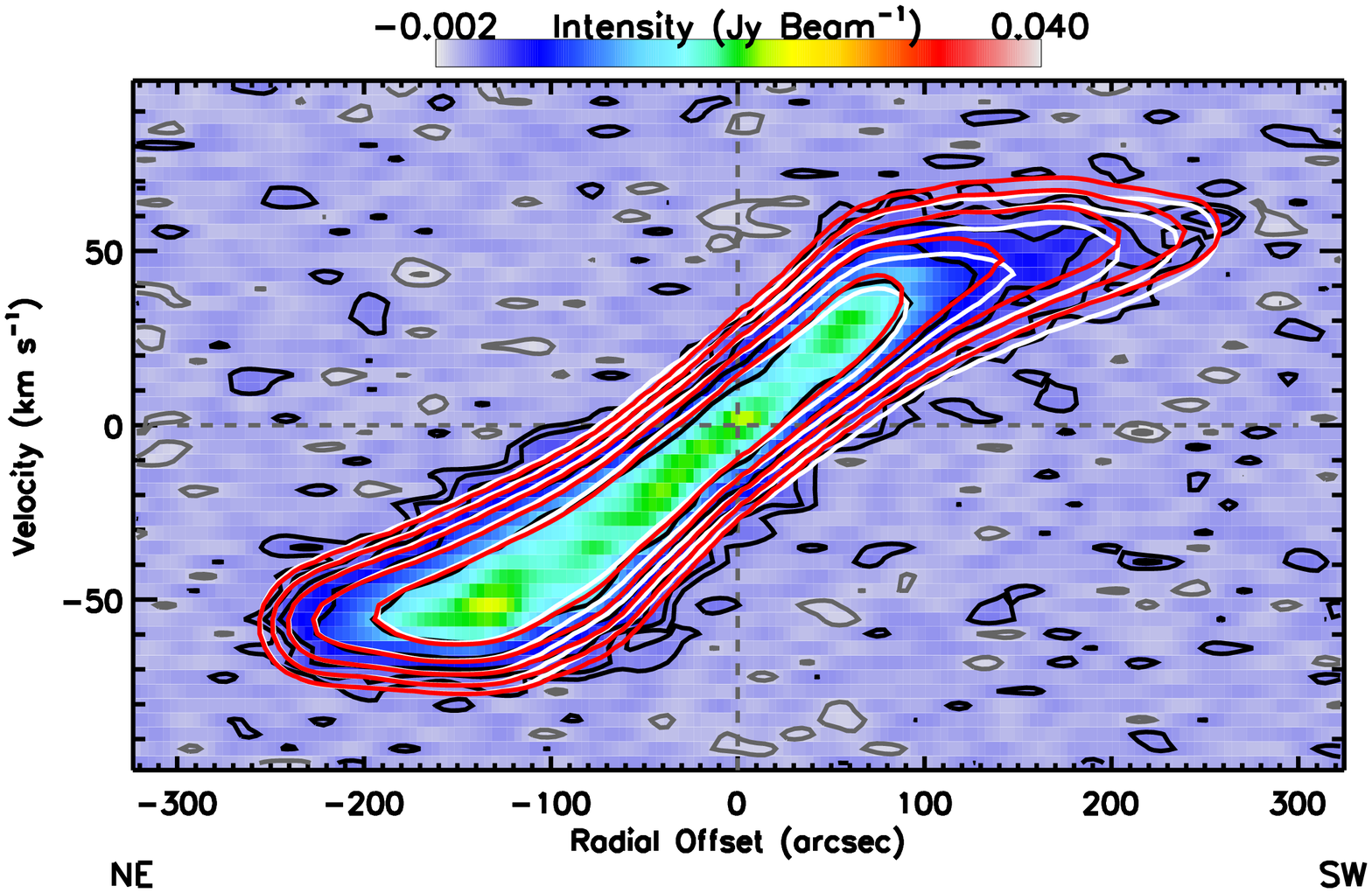}
\includegraphics[angle=0,width=8cm]{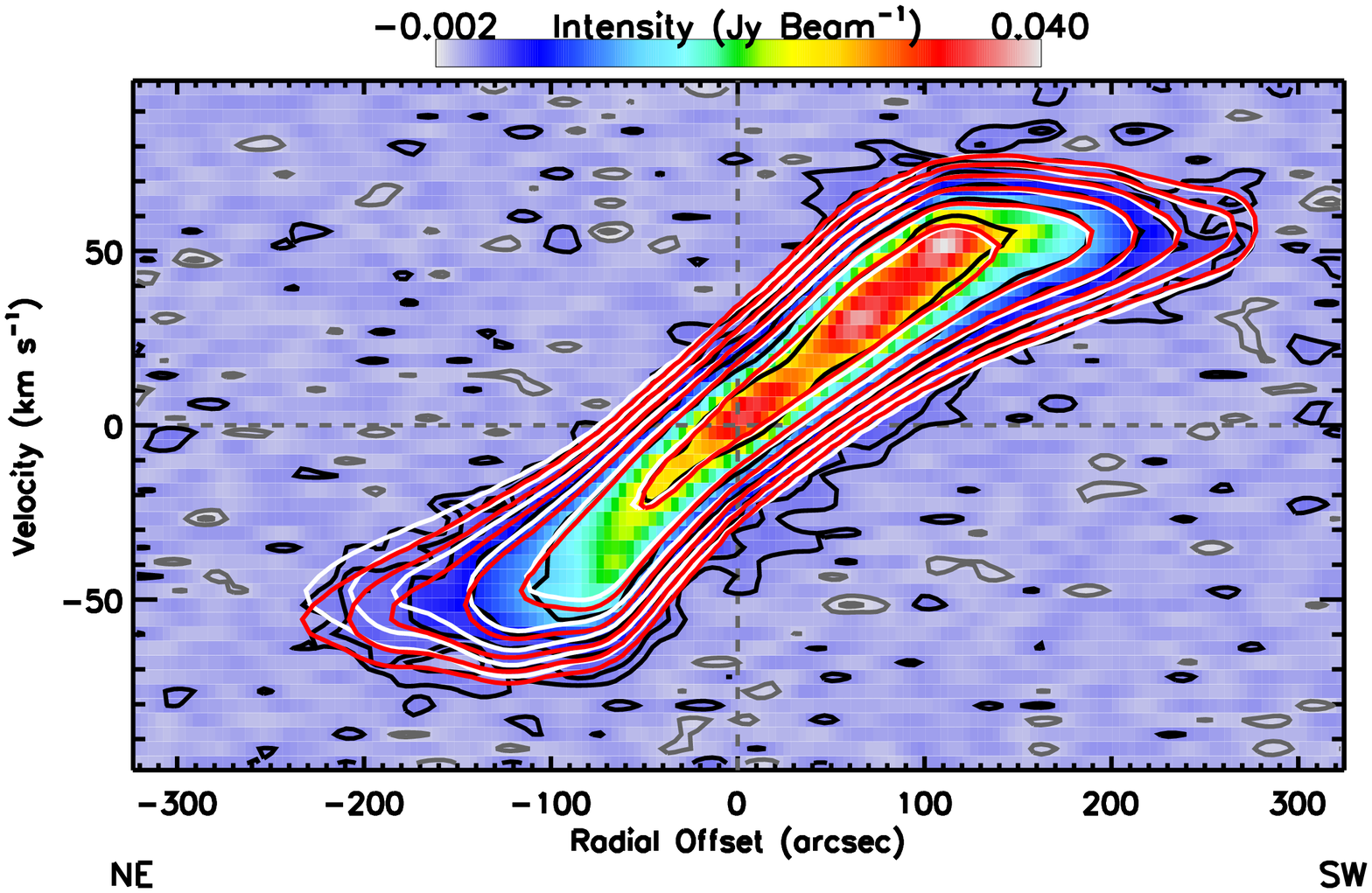}
\caption{Left (Right): Color plot of the {\sc H\,i} PV diagram at 26\arcsec (-26\arcsec)
  ($\pm$0.6 kpc) offset from the major axis. Contours are at -3$\sigma$, - 1.5$\sigma$, 1.5
  $\sigma$, 3 $\sigma$, 6$\sigma$ and so on. The black solid contour is the data, red contour is the  best fit model with no lag, white contour is the best fit model with an assumed vertical gradient of 10.6 km s$^{-1}$  kpc$^{-1}$ in the rotation curve. The colormap ranges are indicated above the panel in Jy/beam. } \label{xv7m6}
\end{figure*}
\begin{figure*}
\centering
\includegraphics[angle=0,width=8cm]{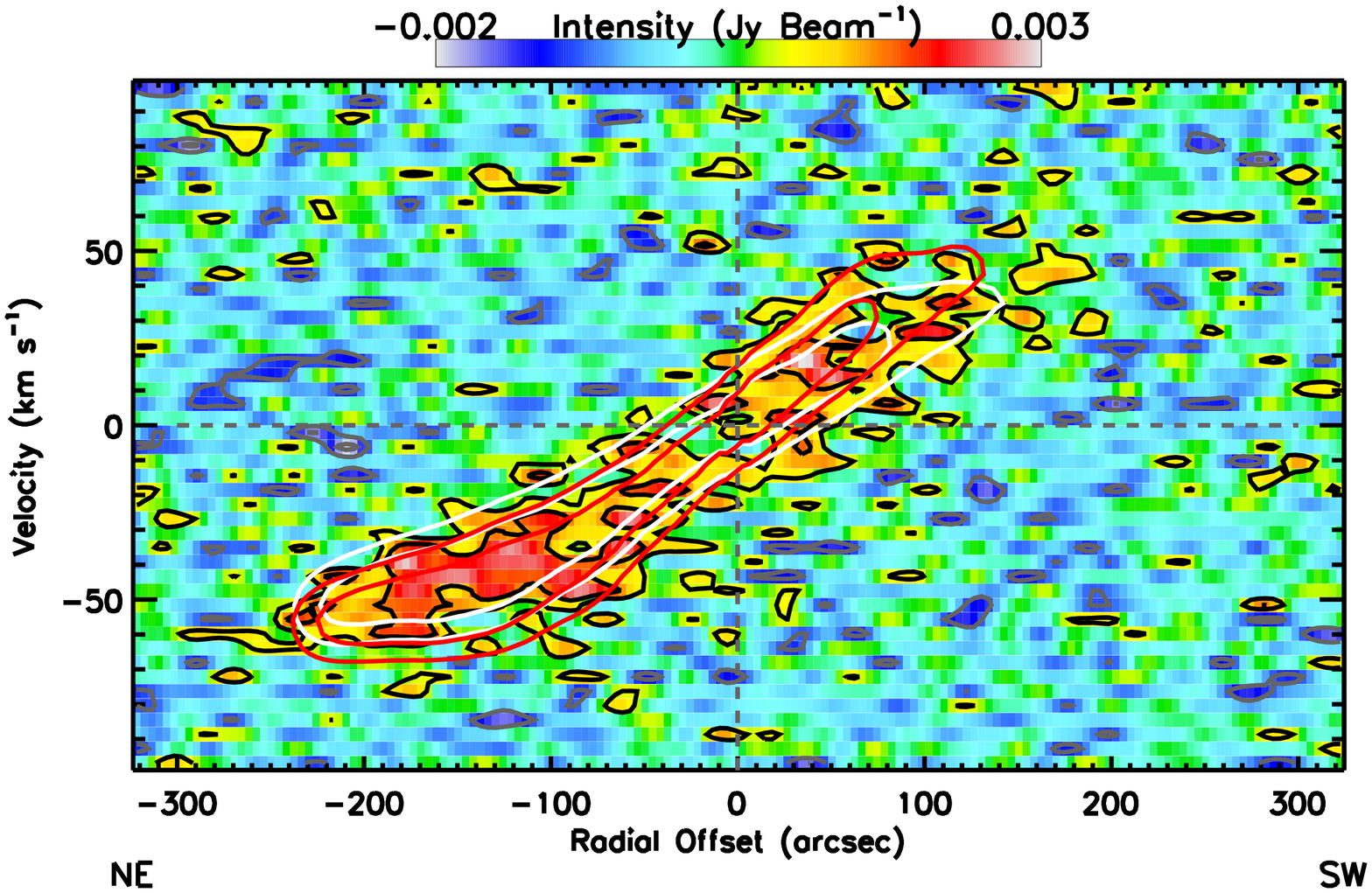}
\includegraphics[angle=0,width=8cm]{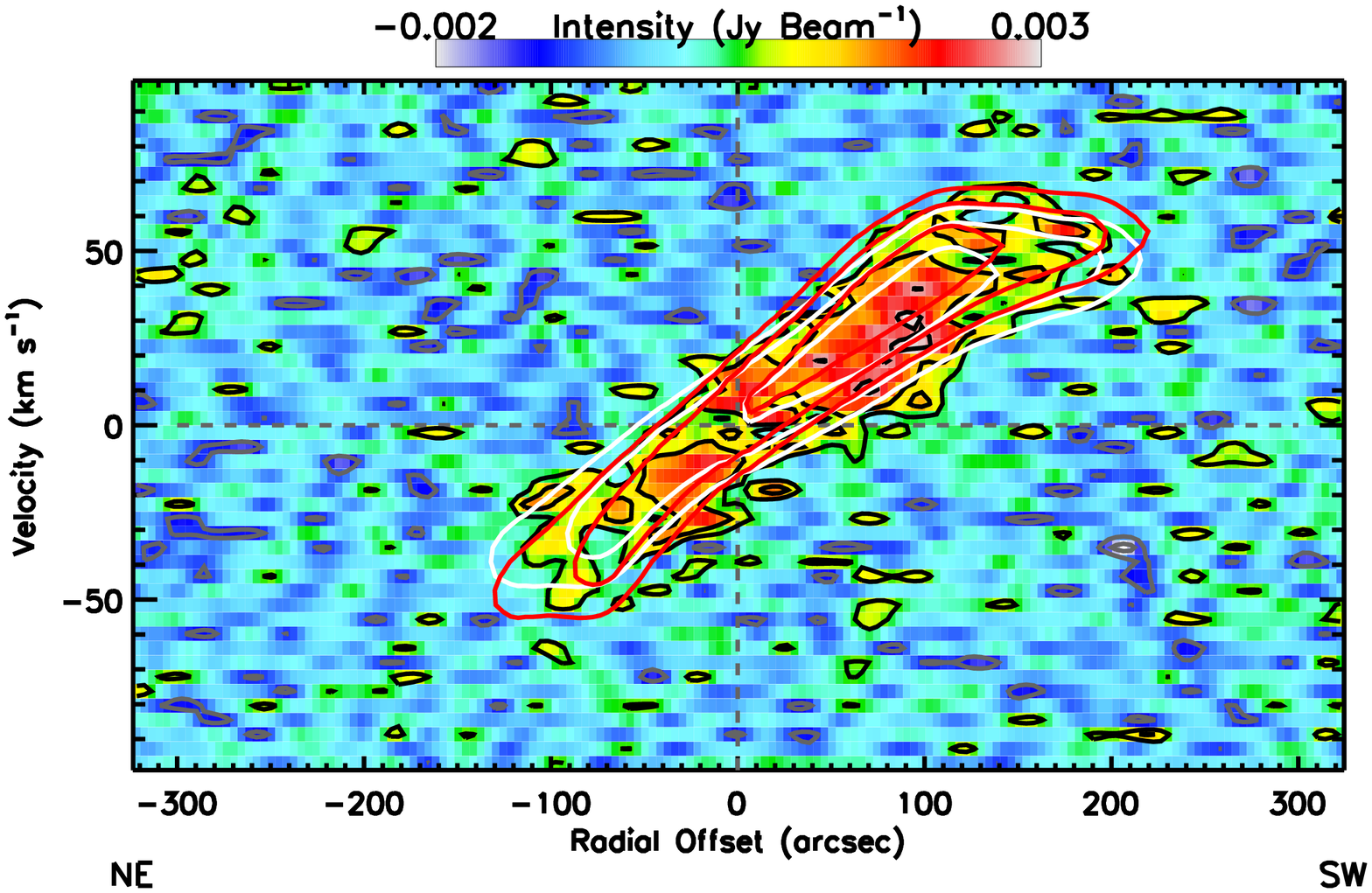}
\caption{Same as Fig. \ref{xv7m6} but now at $\pm$ 56\arcsec ($\pm$1.3 kpc) offset from the plane.  } \label{xv14m13}
\end{figure*}
\hspace*{0.5 cm}As a last check to the model we compare the non-lagging model to the data and the lagging edge-on model in PV diagrams at two distances above (below) the mid-plane. Figures \ref{xv7m6} and \ref{xv14m13}  show PV diagrams parallel to the major axis at $\pm$ 26\arcsec and $\pm$ 56\arcsec offset from the plane.  Already Fig. \ref{xv7m6} is suggesting that  a  model with a lag gives a better fit, though we cannot exclude the non-lagging model at this height. At a height $\sim$60\arcsec above the plane (Fig.  \ref{xv14m13}) it is clearly seen that the velocities around the  3$\sigma$ contours are too high in the non-lagging model.  \\
\hspace*{0.5 cm}Figure \ref{norm} shows also the normalized PV diagrams parallel to the minor axis for the best fit lagging edge-on model. Also in these figures it is now seen that the edge-on model with a vertical gradient and the maximum line-of-sight warp model fit the data equally well. \\ 
\hspace*{0.5 cm} Fig. \ref{HIpara} shows the parameters for the best fit models of UGC 1281 (Edge-on model (black lines) \& Line-of-sight warp model (symbols)). The best fit for the edge-on model has a flare and the scale height ranges from 10\arcsec (0.26 kpc) in the inner parts to a maximum of 15\arcsec (0.39 kpc) on the North Eastern outer parts. The change in PA is the same for both models but in the case of the line-of-sight warp model this is coupled to a change in inclination of the ring as previously explained. The scale length of this model galaxy is 46\arcsec (1.2 kpc). The scale length is the same for the maximum warp model, which is also flaring and has a scale height of 7\arcsec (0.18 kpc) in the inner parts and 12\arcsec (0.31 kpc) in its outer parts. 
\subsection{Ballistic Models}\label{balm}
To investigate what we would expect for UGC 1281 in the sense of extra-planar gas brought up from the disk by supernovae we have used the ballistic model of \cite{2002ApJ...578...98C}.  In this model gas is blown out of the disk into the halo,  with an initial vertical velocity  (called the kick velocity). The model naturally predicts a vertical gradient in rotational velocity which has a higher magnitude in model with a high kick velocity. To obtain the vertical gradient of the model we follow the procedure outlined by \cite{2007ApJ...663..933H}, with the following difference. The initial disk for the model is infinitely thin, this is because the scale height of the H$\alpha$ in UGC 1281 is already comparable to the scale height normally used for the initial disk (0.2 kpc).
Therefore this model should be thought of as an absolute upper limit on the kick velocity.\\
\hspace*{0.5 cm} We find that the scale height of 0.2 kpc is reproduced  when the initial kick velocity is 10 km s$^{-1}$.  We run the simulation until the system reaches steady state ($\sim$1 Gyr). At this point there are no clouds in the halo at radii smaller than 4 kpc due to the radial outward movement of the clouds. Therefore we cannot measure the vertical gradient in the rotational speed at these radii. However, at radii between 5-10 kpc we find a small gradient of 0.4 km s$^{-1}$  kpc$^{-1}$. This is a very low value as one would expect for a small galaxy like UGC 1281 since kick velocities must be small in a low potential, because otherwise the scale height becomes too large.  
 \section{Discussion}\label{disc}
\subsection{Hydrogen in the plane}\label{Hip}
When we compare the parameters of the best fit {\sc H\,i} models to the parameters extracted from the H$\alpha$ data (Table \ref{Hapara}) we see that compared to the ionized gas the neutral gas is much more extended in the radial direction. Apart from this its scale length is slightly larger.  So, the ionized gas is more centrally concentrated than the {\sc H\,i}. \\
\hspace*{0.5 cm} Note that both the neutral gas and the ionized gas display a central depression, or peaks symmetrically around the center, in their distribution (see Section \ref{dist}). Also,  the H$\alpha$ distribution is  more irregular than the neutral hydrogen. This can easily be seen in Figures \ref{mom0Ha} and \ref{mom0HI} by  counting peaks in the distribution (see Section \ref{Hadist}).  One of these 'peaks' in H$\alpha$ has a clear offset from the plane of the galaxy and this region does not stand out  in the {\sc H\,i} observations. \\
\hspace*{0.5 cm}  Another effect is the start of the warp. If we look at the integrated moment maps the neutral gas starts to deviate from the major axis at a radius almost two times larger than  the radius where the H$\alpha$ bends away from the major axis (50\arcsec and 90\arcsec, respectively). Most likely this is a resolution effect, because when we look at the best fit model for the {\sc H\,i} we see both the {\sc H\,i} and the H$\alpha$ start to warp around 50\arcsec (1.3 kpc).\\
\hspace*{0.5 cm} When we compare the measured  velocities of the neutral and ionized gas they look very similar on the major axis (see Fig. \ref{rotcurves}). Ideally this should only occur if the sizes of both disks are similar and there is no absorption, self-shielding or clumpiness affecting the emission.  We know from the analysis of the H$\alpha$  that the velocities in the plane are lower than those above the plane in several places. Also, the {\sc H\,i}  velocities are slightly lower, due to beam smearing, when compared to the best fit model. Since for the H$\alpha$ emission we can only look at  the rising part of the rotation curve, where the rotational velocities are low, deviations from the real rotational velocities are small. Therefore, it is not unreasonable that both effects provide us with the same  velocities since all absolute deviations from the rotational velocities are small. Fig. \ref{rotcurves} (Red line) shows us that the real rotation curve, obtained from the modeling, is slowly rising and reaches a flat part around 60 km s$^{-1}$. This means that UGC 1281 has the slow rising rotation curve which is  common in dwarf galaxies, but it also shows clear differential rotation. \\
\subsection{Hydrogen above the plane}\label{Hap}
We see (Fig. \ref{mom0Ha}) that the H$\alpha$ extends less far in the vertical direction than the {\sc H\,i}. Despite the fact that  a difference in sensitivity may cause us to miss H$\alpha$ emission at distances further from the plane than 20\arcsec, the scale height, measured by fitting a single exponential to the vertical distribution, of the H$\alpha$ is, equal at best, but most likely smaller than the {\sc H\,i} scale height.\\
\hspace*{0.5 cm}Even though the vertical extent of the H$\alpha$ is limited and not supplying much information above the plane the {\sc H\,i} extends well above the plane and much can be learned from it. If we assume a constant gas dispersion of 9 km s$^{-1}$ we would expect a scale height of $\sim$400 pc for the disk, from a simple estimate \citep{1992AJ....103.1841P}. This implies that the gas at large distances from the mid-plane is not an additional component but merely the high-latitude gas of a pressure supported disk.  This idea is strengthened by the fact that a ballistic model requires a very low kick-velocity ($\sim$10 km s$^{-1}$), of the order of the velocity dispersion in the model, and indicates that the gas at large distances from the mid-plane is not pushed to high latitude by star formation processes. Furthermore, continuum maps constructed from the line-free channels of the 21 cm observations do not show any continuum emission at the position of UGC 1281 once more confirming the low SFR. This implies that UGC 1281 does not contain a typical halo such as seen in NGC 891. However, also in the case of UGC 1281 the kinematics show that as the distance from the mid-plane increases the projected mean velocities decline (see Fig. \ref{norm}). \\ 
\hspace*{0.5 cm}From our modeling we obtain two possible explanations for the high latitude {\sc H\,i} and its declining projected mean velocities. We find that a maximum line-of-sight warp model, where we push the warp as much into the line of sight as the data allows, fits the data equally well as a purely edge-on model with {a warp in the plane of the sky and }a vertical gradient in its rotation curve of -10.6 $\pm$ 3.7 km s$^{-1}$ kpc$^{-1}$. Both models assume a single exponential in the vertical direction and the edge-on model does assume that the radial gas distribution above the plane is similar to that in the disk. In theory the lower velocities could also be caused by an alternate distribution above the disk. However, since the vertical gradient is seen almost everywhere in the observations, it seems unlikely that this is the case unless  the gas is in a warp or flare and these possibilities are included in the modeling.\\
\hspace*{0.5 cm}To outline the importance of separating between a lag and a line-of-sight warp we will discuss these two options and their implications in two separate sections below. These discussions are by no means meant as a complete overview of the theory and observations behind extra-planar gas but only to highlight the parts where UGC 1281 can significantly contribute to a better understanding of the theories. For a full and complete review of cold extra-planar gas we refer the reader to \cite{2008A&ARv..15..189S} and references therein. \\ 
\subsubsection{A warp}
The first possibility that can explain the observations is that the gas at large projected distances from the mid-plane in UGC 1281 is located in a simple warp. Warps are quite common in disk galaxies and often they are asymmetric \citep{2002A&A...394..769G} as is observed in UGC 1281. If the {\sc H\,i} and its kinematics in UGC 1281 are to be fully explained by a warp, the warping axis has to have an angle of 55$^{\circ}$ with respect to the line of sight.\\
\hspace*{0.5 cm}We would like to point out the very small differences between the edge-on model and the maximum line-of-sight warp model.  The maximum difference in inclination between the two models is 10$^{\circ}$ but on average 5$^{\circ}$ in inclination. This results in a difference in scale height of $\sim$3\arcsec (see Fig. \ref{HIpara}). These differences take away the need for a vertical gradient completely. This once more shows that differences between line-of-sight warps and  lagging halos are very subtle, and that great care must be taken to exclude one of the two options.\\
\hspace*{0.5 cm}When we measure once more the maxima in normalized PV diagrams parallel to the minor axis we find an apparent vertical gradient of  4.9 $\pm$ 1.1 km s$^{-1}$ kpc$^{-1}$, for the maximum line-of-sight warp model, which is consistent with the measurement from the data (6.9 $\pm$ 2.0 km s$^{-1}$ kpc$^{-1}$). Therefore there is no need to introduce a lag or other extra kinematical effects  into this model.\\
\hspace*{0.5 cm}The best fit model  for this case has central scale heights which are similar to  those measured from the stars and the H$\alpha$. This would mean that the stars and the ionized gas hardly extend into the warped outer regions of the disk and that all the  {\sc H\,i} at large projected distances from the mid-plane is in the warp and flare. We have already seen that the PA starts changing well within the maximum radius of the H$\alpha$ as well as inside the optical radius. This is seen for every model, whether it is  edge-on or maximum line-of-sight. \\
\hspace*{0.5 cm}The start of the warp could be affected by a slight error in the assumed PA. We have tested this by rotating the integrated {\sc H\,i} moment map  and the red DSS image (See Fig. \ref{mom0HI}) by an additional 1 and 2 degrees (PA= 39$^\circ$, 38${^\circ}$ respectively) and plotting the vertical offsets of the peak of the vertical profiles. This shows that for the red optical image the PA of 40$^\circ$ gives the flattest central distribution but that the start of the warp in the  {\sc H\,i} can be pushed outward by $\sim$ 15\arcsec by assuming a PA of  39$^\circ$.  This means that the observed start of the warp remains well within the optical radius ($\frac{1}{2}$D$_{25}$=134\arcsec) even if the assumed PA is slightly off.\\
 \hspace*{0.5 cm}The fact that the warp starts within the optical radius is inconsistent with the findings of \cite{2007A&A...466..883V} that warps start just beyond the truncation radius of the stellar disk. Even more so, we know from the observed H$\alpha$ emission that also the ionized gas is slightly warped (see $\S$ \ref{Hadist}). This would imply that if the warp in UGC 1281 is formed by the accretion of gas from the IGM its initial disk was not rigid enough to stabilize the infalling gas \citep{2007A&A...466..883V}. If so, one would expect a clear difference between the onset of the warp, with respect to its truncation radius, between dwarf galaxies and massive galaxies. \\
 \subsubsection{A lag}\label{laghal}
In the case of a warp which is purely in the plane of the sky the modeling indicates a larger scale height of the {\sc H\,i} than the scale heights measured from the stars and the ionized hydrogen. This vertical density distribution can be modeled with a single exponential, however the kinematics indicate the need for a vertical gradient in the rotation curve when the warp is purely perpendicular to the line of sight. As in the case of the super-thin galaxy UGC 7321 \citep{2003ApJ...593..721M} the origin of this lagging neutral hydrogen gas in a Low Surface Brightness (LSB)  dwarf galaxy, with low star formation rates, would be puzzling. \\
\hspace*{0.5 cm}If the high latitude gas in UGC 1281 is lagging this would have some implications to current theory. \cite{2007ApJ...663..933H} have compared the lag, or the vertical gradient in the rotation curve in three massive galaxies. They find tentative evidence that when they scale the lag with the observed H$\alpha$ scale height\footnote{Note that \cite{ 2007ApJ...663..933H} use electron scale heights instead of the emission measure scale heights.} this new parameter ($dV/dh_z$) is roughly constant at about 20 km s$^{-1}$ scale height$^{-1}$. However, the galaxies compared are of similar mass. The results presented in this paper give us now the opportunity to take a tentative look at a class of galaxies with much lower mass. The dynamical mass of UGC 1281 is 6.3$\times10^9 \rm{M}_{\odot}$, measured at our last point of the rotation curve, as opposed to $\sim 1\times10^{11} \rm{M}_{\odot}$ for the galaxies in the study by \cite{2007ApJ...663..933H}.\\
\hspace*{0.5 cm}Within the picture of a lagging disk, we would have $dV/dh_z$ = 4.7 $\pm$ 1.7 km s$^{-1}$ per scale height. This value would be inconsistent with a constant $dV/dh_z$ $\sim$20 km s$^{-1}$ which was found for the 3 massive galaxies. As shown by the ballistic models ($\S$ \ref{balm}), the lag expected purely on gravitational grounds would be much shallower than the one than the one that we include in this model. Such a lag would thus require additional effect to be at play such as described by e.g. \cite{2002ASPC..276..201B,2006A&A...446...61B,2008MNRAS.386..935F}. In any case,  star formation in the disk would be of negligible influence on the vertical gradient.\\
\hspace*{0.5 cm}The previous discussion clearly shows the need  to unambiguously determine a lag in a small galaxy such as UGC1281 as well as the need for larger sample of galaxies with quantified vertical velocity gradients. In the present case, due to the fact that the addition of a lag to the model of UGC 1281 does not significantly improve the match with the data, we prefer the conceptually simpler line-of-sight warp model. \\
\section{Summary}\label{summary}
We presented  21 cm and H$\alpha$ emission in the edge-on dwarf galaxy UGC 1281. This is the first time such sensitive {\sc H\,i} data have been presented for a dwarf edge-on.\\
\hspace*{0.5 cm} The integrated H$\alpha$ velocity map (Fig. \ref{mom0Ha}) shows a non-smooth distribution on the major axis with several peaks. One of these peaks is actually located beneath the major axis. It is unclear whether this {\sc H\,i\,i} region is  located above the plane of the galaxy or in its warped outer parts.\\
\hspace*{0.5 cm} The integrated {\sc H\,i} velocity map (Fig. \ref{mom0HI}) shows a quite regular distribution with a central depression. This central depression appears to be symmetrical in position around the center but from modeling it follows that it is somewhat deeper on the NE side of the galaxy. Such a central depression is not  uncommon for dwarf galaxies. \\
\hspace*{0.5cm}Furthermore this map shows that UGC 1281 is warped  in its outer parts and this warp resembles a ``normal" S-shape at its start. However, at large radii the warp bends back towards the inner plane of the galaxy.\\
\hspace*{0.5 cm}For the interpretation of the kinematics of the
high latitude {\sc H\,i} gas we constructed velocity maps from the H$\alpha$ and {\sc H\,i} data. Also  3-D models with a modified version of GALMOD are constructed in GIPSY. This modified version enables us to construct kinematic models with a vertical gradient. \\
\hspace*{0.5 cm} The velocities obtained from the data show a slow rise in the inner part. This is also seen in the rotation curve obtained from the modeling and therefore unlikely to be an effect of the {\sc H\,i} distribution or resolution. At about 120\arcsec\hspace{0.1 cm}the rotation curve flattens off to a maximum rotational velocity $\sim 60$ km s$^{-1}$. This slow rise is common for dwarf galaxies.\\
\hspace*{0.5 cm} From our modeling we find that our data are not sensitive enough to distinguish between a lag or a line-of-sight warp. Both models fit the data equally well and there is only a small difference between the input parameters of the models. However, the models do start to deviate at emission levels slightly lower than our current sensitivity limit. Therefore one would most likely be able to separate between the two models with deeper observations. \\
\hspace*{0.5 cm} In the case of  lagging high latitude gas the low vertical extent and the low flux level of the H$\alpha$ emission would indicate that this high latitude {\sc H\,i} does not originate from galactic fountains. The {\sc H\,i} scale height implies a normal pressure supported disk and thus no need beyond turbulence for a mechanism to bring it up from the mid-plane. If in such a  disk  the pressure, of the gas, is solely dependent on the density (e.g. barytropic) one would expect it to be co-rotating. However, we find in our analysis that the high latitude gas in this case has a lag of 10.6 $\pm$ 3.7 km s$^{-1}$ kpc$^{-1}$ when compared to tilted ring models. This lag could be caused by infalling gas or pressure gradients above the mid-plane. \\
\hspace*{0.5 cm} In the case of a line-of-sight warp the ionized hydrogen and the distribution of the stars would, for the most part, not extend into the warped region of the disk. However, the scale height in the central parts would be the same for the stars, H$\alpha$ and {\sc H\,i}.\\
\hspace*{0.5 cm}Regardless of  which model fits the data best, maximum line-of-sight warp or edge-on, the warp starts well with the optical radius (D$_{25}$=4.46\arcmin, \cite{1992yCat.7137....0D}), at a radius of $\sim$ 50\arcsec, which is unlike more massive galaxies \citep{2007A&A...466..883V}.\\
\hspace*{0.5 cm}The small differences in input parameters between a model with a lag and one with a line-of sight warp show that great care must be taken to distinguish between lagging halos and line-of-sight warps since small changes in the modeling can have a great effect on the velocity field.\\
\hspace*{0.5 cm} Summarising our main conclusions:
\begin{itemize}
\item{The rotation curve of UGC 1281 is slowly rising in its inner parts and flattens off to a maximum rotational velocity $\sim$60 km s$^{-1}$ at 120\arcsec (3.14 kpc). }
\item{The neutral hydrogen in UGC 1281 is more extended than the stars and the ionized emission in its radial distribution.}
\item{The gaseous warp start well within the optical radius.}
\item{Our observations can be fitted by both a vertical gradient in the rotation curve and a line-of-sight warp. The observations are not sensitive enough to separate between these two options. However, the line-of-sight warp model is conceptually simpler and therefore preferable.}
\end{itemize}
\section*{Acknowledgements}
We thank R. Kuzio de Naray for making her velocity data available to us. We are grateful to S. Sanchez and A. Guijarro for doing the PPAK observations and M. Verheijen for introducing us to PPAK. Many thanks to T. Oosterloo for help with the {\sc H\,i} data reduction and to P. Serra, R.-J. Dettmar, D.J Bomans and our anonymous referees for useful comments, discussion and insights. 
\bibliographystyle{mn2e}   \bibliography{ref}

\bsp

\label{lastpage}

\end{document}